\documentclass[]{aa}
\usepackage{lscape,graphicx}
\usepackage{natbib}
\bibpunct{(}{)}{;}{a}{}{ }
\begin{document}

\title{ On the artificial nature of aperiodic variability in XMM-Newton observations of M31 X-ray sources  and the ultraluminous X-ray source NGC 4559 ULX-7}
\author{ R. Barnard\inst{1}, S. Trudolyubov\inst{2}, U. C. Kolb\inst{1}, C. A. Haswell\inst{1}, J. P. Osborne\inst{3}, W. C. Priedhorsky\inst{4}
}

\offprints{R. Barnard, \email{r.barnard@open.ac.uk}}

\institute{ The Department of Physics and Astronomy, The Open University, Walton Hall, Milton Keynes, MK7 6BT, U.K.
      \and  Institute of Geophysics and Planetary Physics, University of California, Riverside, CA 92521
     \and The Department of Physics and Astronomy, The University of Leicester, Leicester, LE1 7RH, U.K.
 \and Los Alamos National Laboratory, Los Alamos, NM 87545}

\date{ Received 01/09/2006 / Accepted 05/03/2007}

%\abstract{}{}{}{}{}
\abstract {Power density spectra (PDS) that are characteristic of low mass X-ray binaries (LMXBs) have been previously reported for M31 X-ray sources, observed by XMM-Newton. However, we have recently discovered that these PDS result from the improper addition/subtraction of non-simultaneous lightcurves.}
{ To understand the properties and origins of the artefact.}
{ We re-analysed our XMM-Newton observations of M31 with non-simultaneous and simultaneous lightcurves, then combined simulated lightcurves at various intensities with various offsets and found that the artefact is more dependent on the offset than the intensity.}
{ The lightcurves produced by the XMM-Newton Science Analysis Software (SAS)  are non-synchronised by default. This affects not only the combination of lightcurves from the three EPIC detectors (MOS1, MOS2 and pn), but also background subtraction in the same CCD. It is therefore imperative that all SAS-generated lightcurves are synchronised by time filtering, even if the whole observation is to be used.  We also find that the  reported timing behaviour for NGC 4559 ULX-7 was also contaminated by the artefact; there is no significant variability in the correctly-combined lightcurves of NGC 4559 ULX-7. Hence, the classification of this source as an intermediate-mass black hole is no longer justified.} 
{\rm  While previous timing results from M31 have been proven wrong, and also the broken power law PDS in NGC 4559 ULX-7, XMM-Newton was able to detect aperiodic variability in just 3 ks of observations of NGC 5408 ULX1. Hence XMM-Newton remains a viable tool for analysing variability in extra-galactic X-ray sources. 
}

\authorrunning{Barnard et al.}
\titlerunning{ Artificial stochastic variability in combined EPIC lightcurves}
\maketitle

\section{Introduction}

The variability and spectral properties of Galactic low mass X-ray binaries (LMXBs) are well known to depend more on the accretion rate than on whether the primary is a neutron star or black hole \citep{vdk94}. At low accretion rates, the power density spectra (PDS) may be characterised by broken power laws, with the spectral index, $\gamma$, changing from $\sim$0 to $\sim$1 at some break frequency in the range 0.01--1 Hz \citep{vdk94}; such variability has a r.m.s power of $\sim$10--40\% \citep[e.g.][]{vdk95}.  We describe such PDS as Type A \citep{bko04}. At higher accretion rates, the PDS is described by a simple power law with $\gamma$ $\sim$1 and r.m.s. variability $<$10\% \citep{vdk94,vdk95}. We refer to these PDS as Type B \citep{bko04}.

Type A variability is characteristic of disc-accreting X-ray binaries; any X-ray source that exhibits such variability may be identified as an X-ray binary, rather than a foreground star or background active galaxy \citep[see e.g.][]{bok03}. Such variability has been reported in  XMM-Newton observations of M31 by some of the authors \citep[see e.g.][]{bok03, bko04,will05}. Type A PDS have also been reported for ultraluminous X-ray sources (ULXs) in NGC\thinspace 4559 \citep{crop04}  and NGC\thinspace 5408 \citep{sor04}. 

 However, we have now discovered an  artefact that produces false Type A PDS, caused by the improper addition of lightcurves from the three EPIC instruments on board XMM-Newton (MOS1, MOS2 and pn), where the lightcurves are not synchronised (Barnard et al., this issue). The data were analysed using SAS 6.0.0 and FTOOLS 5.3.1.
In the Sect.~\ref{man} we  discuss how the error occurred. We then describe a new, corrected analysis of the M31 data in Sect.~\ref{newanal}. Section \ref{sims} describes our investigation into the  artefact using simulated lightcurves. Then, Sect.~5 details the causes of the artefact. Section \ref{ulx} re-analyses XMM-Newton observations of two ultraluminous X-ray sources with published Type A PDS.  We draw our conclusions in Sect.~\ref{conc}.

\begin{figure*}[!t]
\resizebox{\hsize}{!}{\includegraphics[angle=270]{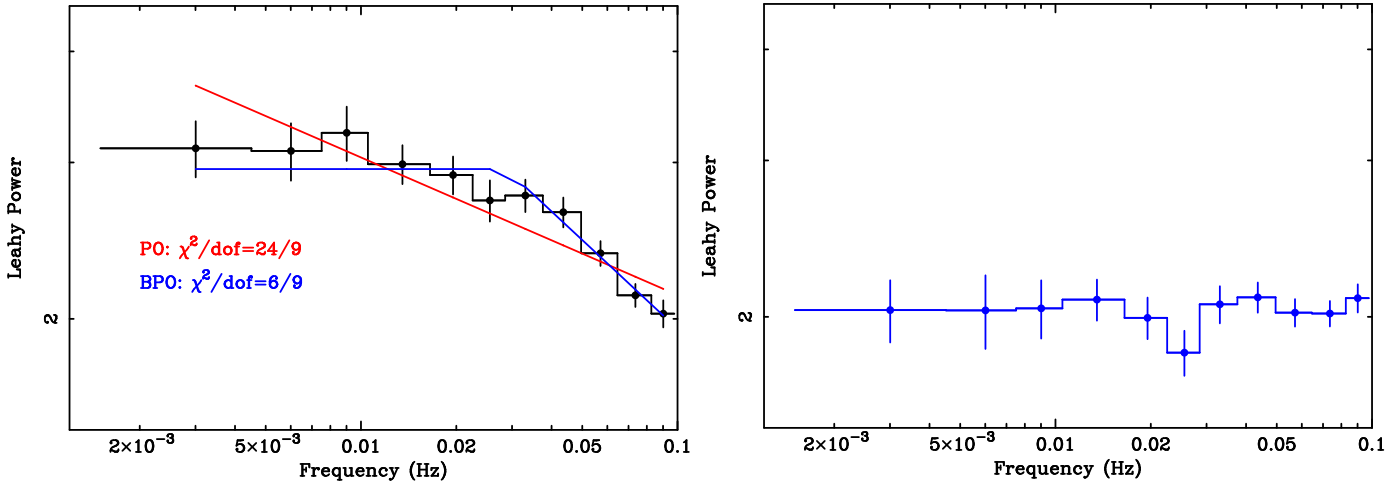}}
\caption{ Power density spectra of 0.3--10 keV lightcurves of r3-60, using Method I (left) and Method II (right). We present best fits to the Method I PDS of power law and broken power law models; a broken power law is clearly required. The PDS is hence reminiscent of  Type A variability in  Galactic LMXBs. However, the Method II PDS is flat, revealing that this variability is artificial.
}\label{fig4}
\end{figure*}

\section{Manipulating lightcurves generated by SAS} 
\label{man}
XMM-Newton has observed M31 extensively, and \citet{bok03, bko03, bko04,bfh06}  conducted a survey of bright sources in the central region, over four observations. For each observation of every source, 0.3--10 keV  lightcurves were obtained  from source and background regions in the  MOS1, MOS2 and pn images,   using 2.6 second binning. 

We have recently discovered that the results of manipulating lightcurves produced with the SAS tool {\bf evselect} using the  {\bf lcmath} FTOOL produces different results, depending on how the time selection in {\bf evselect} is phrased. 
Originally, lightcurves  were produced from events files that were filtered in {\bf evselect}  with the expression ``{\em (PI in [300:10000]) \&\& (TIME in [{\it tstart}:{\it tstop}]}''; additional filtering was instrument dependent: MOS lightcurves were additionally filtered using ``{\em $\#$XMMEA\_EM\&\&(PATTERN$<$=12)}'', while the pn equivalent was ``{\em$\#$XMMEA\_EP\&\&(PATTERN$<$=4)}''. We shall refer to this filtering as Method I.

However, different lightcurves are obtained when {\bf evselect} is used in either of the following ways.   Firstly, one may assign {\em tstart} and {\em tstop} to  the TLMIN1 and TLMAX1 keywords in the pn, MOS1 and MOS2 events files, using FTOOLs such as {\bf fmodhead} or {\bf fparkey}; we refer to this method as Method II.  Method III involves   filtering each lightcurve with the {\bf evselect} tool, using the parameters ``timemin={\it tstart} timemax={\it tstop}''.  Methods II and III are equivalent. Vitally, they are not equivalent to Method I, although this is not mentioned in any documentation.  Consequently, we analysed lightcurves obtained with each method to determine which, if any, were correct.

\section{Re-analysing the data}
\label{newanal}

\subsection{Combined EPIC lightcurves}

 We obtained MOS1, MOS2 and pn lightcurves from the source extraction regions of several sources in the M31 observations, using Methods I and II. 
 For Method I, we found that uncertainties of  combined EPIC (MOS1+MOS2+pn)  lightcurves were underestimated by a factor of $\sim$10--40\%. This ratio varied within each lightcurve, with the standard deviation in the ratio for a given lightcurve decreasing with increasing luminosity. The reason for the small uncertainties is discussed in Sect. 5.1.

Closer inspection of the lightcurves has revealed the difference between Method I and Method II lightcurves: the Method II lightcurves are synchronised, while the Method I lightcurves are not.  Of course, for the Method II or III lightcurves to be synchronised, {\em tstart} must be the same for all detectors, and must be present in all the events lists.

Many PDS from  lightcurves combined using Method I exhibit Type A variability, while identical data combined with Method II do not.  To investigate the cause of the Type A variability in the observed PDS, we examined the Method I and II lightcurves of XMMU\thinspace J004208.9+412048 \citep[r3-60][]{K02}. The Method I and Method II PDS are shown in the left and right panels respectively of Fig.~\ref{fig4}; they were obtained using the FTOOL {\bf powspec}.  The Method I PDS is well modelled ($\chi^2$/dof = 6/8) by a broken power law where  the spectral index  changes from 0 to 0.33$\pm$0.05 at a frequency of 29$\pm$4 mHz. Such a PDS is characteristic of Galactic LMXBs. However, the PDS of the Method II lightcurve is flat, with a Leahy power of 2, consistent with Poisson noise. Hence the variability in the  Method I PDS is artificial.

.   
\begin{figure}[!t]
\resizebox{\hsize}{!}{\includegraphics[angle=270]{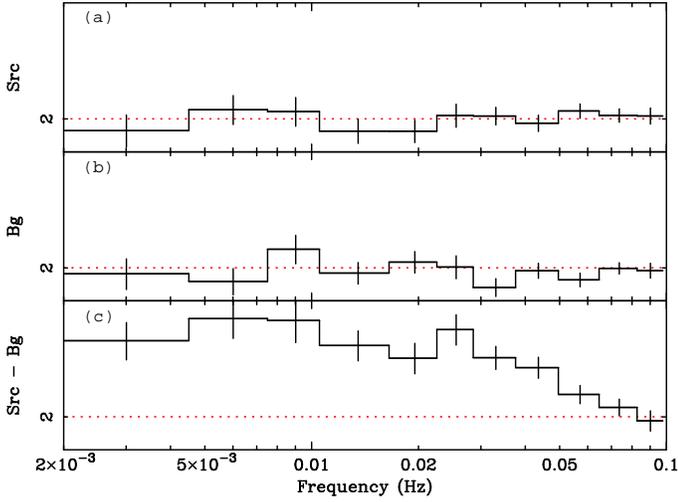}}
\caption{  PDS from source+background (a), background (b) and background-subtracted (c) pn Method I lightcurves of r3-125. The background-subtracted PDS is of Type A, although the component PDS are flat. 
}\label{fig6}
\end{figure}

\subsection{Background Subtraction}

Background subtraction  requires two lightcurves: source and background. We produced background-subtracted pn lightcurves of various X-ray sources from the 2002 January XMM-Newton observation of the M31 core, to test whether the lightcurves from the same detector were synchronised.  We present the Method I PDS from source, background and background subtracted pn lightcurves of r3-125 in Fig.~\ref{fig6}, panels (a), (b) and (c) respectively. The source and background PDS are flat, while the background-subtracted PDS is of Type A. Hence, background subtraction causes artefacts when  non-synchronised source and background lightcurves are used. 

\section{Investigating the artificial variability with simulated lightcurves}
\label{sims}

To determine the cause of artificial variability we combined artificial lightcurves with varying intensities and offsets. We defined 100 evenly spaced intensity intervals from 0.01 to 1.00 count s$^{-1}$, and for each interval generated a single ``source'' lightcurve and 100 ``background'' lightcurves (using the terminology of {\bf lcmath}:   the ``background'' lightcurve may be added to or subtracted from the ``source'' lightcurve). This intensity range reproduces well the observed intensities from the bright X-ray sources in the M31 Core. Each lightcurve had  60 ks lightcurve duration, with 2.6 s binning.

 For each lightcurve, we generated a series of photons with random arrival times, using the Park \& Miller Minimal Standard random number generator, shuffled by the Bays \& Durum method (Knuth, 1981); the random number seed for the $n$th ``background'' lightcurve at intensity $i$ was defined as $-(10000i+n)$.  The source lightcurves were generated with seeds of $-10000i$;  hence, each lightcurve was reproducible and unique.
We then conducted a series of tests for artificial power  in the PDS of combined lightcurves.

\subsection{Adding two lightcurves with different start times}
Firstly we added each ``background'' lightcurve for every  intensity interval to the corresponding ``source'' lightcurve at various offsets with {\bf lcmath}; the offset was defined as the delay in seconds as a fraction of the 2.6 s time binning, and ranged from 0.05 to 0.95 at intervals of 0.05. We then produced  geometrically grouped, 64-bin PDS of the resulting lightcurve with  the FTOOL {\bf powspec}, averaged over $\sim$180  time intervals. Initially, we examined the PDS for excess variability  by fitting them with  a  constant Leahy power of 2; for our PDS, a $\chi^2$ $\ga$25 / 10 dof  signified a 3$\sigma$ detection of power above the Poisson noise.  We find that the offset is more important than intensity, with all combinations of lightcurves with offsets 0.25--0.80 producing significant excess variability;   12\% of lightcurves with an offset of 0.05 or 0.95 produced artificially variable PDS. However, not all variable PDS are considered Type A. 

To be classed as Type A, a power law fit to the PDS must be rejected, and a broken power law fit must be reasonable. A broken power law fit may be described by spectral index $\alpha$ that changes to spectral index $\beta$ at frequencies higher than  some break frequency $\nu_{\rm b}$; the final parameter is the power at $\nu_{\rm b}$. When fitting the PDS, we rejected power law fits with good fit probability $<$0.005 ($\chi^2$ $>$24) and accepted broken power law fits with good fit probability $>$ 0.05 ($\chi^2$ $<$ 16); for the broken power law fit, we set $\alpha$=0, as observed in Galactic LMXBs.  Less than 10\% of lightcurves with offsets in  the ranges 0.05--0.25 and 0.75--0.95 exhibited Type A variability, while 30\% of lightcurves with $f=0.50$ were of Type A.

\begin{table}[!b]
\caption{ Summary of variability observed when three artificial lightcurves are combined, for Cases A and B. For each case, the fraction of combined lightcurves exhibiting Type A were shown, for offsets $f=0.05$, $f=0.50$ and $f=0.95$. }
\label{offper}
\begin{tabular}{cccc}
\noalign{\smallskip}
\hline
\noalign{\smallskip}
Case & $f=0.05$ & $f=0.50$ & $f=0.95$\\
\noalign{\smallskip}
\hline
\noalign{\smallskip}
A & 0.003 & 0.13 & 0.011 \\
B & 0.12 & 0.75 & 0.12 \\
\noalign{\smallskip}
\hline
\noalign{\smallskip}
\end{tabular}
\end{table}

\subsection{ Adding three lightcurves with different start times}
\label{3lc}
Since Barnard et al. combined source lightcurves from the three EPIC detectors in the original M31 survey work \citep[e.g.][]{bok03,bko03,bko04} we added a third lightcurve to the combined artificial lightcurves discussed in the previous section.  For Case A, this lightcurve was the 50th background lightcurve with an offset of 0.1, and for Case B  the 50th background lightcurve with an offset of 0.5 was added.  The resulting lightcurves had intensities of 0.03--3 count s$^{-1}$, with offsets of 0.05--0.95 as before. 

 We  tested the resulting PDS for Type A variability. The results are summarised in Table~\ref{offper}. Again, we found that the fraction of Type A lightcurves dependent on offset rather than intensity, with the maximum at $f$ = 0.50.  However, even an offset of 0.05 produced artificial Type A variability in 12\% of combinations for Case B.  Combining three non-synchronous lightcurves (in Case B) results in a factor of $>$2 more false Type A variability than just combining two lightcurves. Hence the artefact is strengthened as more non-synchronous lightcurves are added (combining detectors) or subtracted (background subtraction).

\section{ The causes of the artefact}

Three factors contribute to the artificial Type A variability: underestimated uncertainties, excess power in the PDS, and suppression of high-frequency variability. We now discuss these in turn.
\subsection{ Underestimating Method I uncertainties}

 The smaller uncertainties of Method I are due to the workings of {\bf lcmath}. Whether adding or subtracting lightcurves, {\bf lcmath} treats the first lightcurve as  the source and the second lightcurve as the background; it calculates the mean ``background'' intensity for each time interval in the ``source'' lightcurve. If the lightcurves are offset by some non-integral number of time bins, then the intensities of two ``background'' intervals are weighted by their overlap with the ``source'' interval. The intensity of the $n^{\rm th}$ interval of the summed lightcurve, $S\left( n\right)$, is then given by
\begin{equation}
\label{sumcalc}
S\left( n\right) = A\left( n\right) +f\times B\left( n-1\right)
+ \left( 1 - f\right)\times B\left( n\right) 
\end{equation}
where $A$ is the ``source'' lightcurve, $B$ is the ``background'' lightcurve and $f$ is the fractional offset. The latter is  given by
\begin{equation}
f =  \frac{t_{\rm 0}^{\rm B} - t_0^{\rm A}}{\Delta T}
\end{equation}
where $t_{\rm 0}^{\rm A}$ and $t_{\rm 0}^{\rm B}$ are the start times for the source and background lightcurves, respectively, and $\Delta$T is the width of the time interval. 

 By default, the lightcurves begin at the arrival time for their first photons. In practise, however, $t_0^{\rm A}$ is determined by the arrival of the first photon in the ``source'' lightcurve but {\bf lcmath} sets $t_{\rm 0}^{\rm B}$ to the start of the first  ``background'' interval to overlap the ``source'' lightcurve. In addition, {\bf lcmath} ignores all source intervals with no ``background'' interval. Hence, 0 $\le$ $f$ $\le$ 1. For synchronised lightcurves, $t_0^{\rm A}$ and $t_0^{\rm B}$ are both set to the same user-defined value. Hence $f$ = 0 and $S\left( n\right) = A\left( n\right) + B\left( n\right)$,  as it should be.

The underestimated uncertainties are caused by an error in the {\bf lcmath} code {\bf lcmathrdbkg.f}. When two or more background bins coincide with a source interval, with count rate $C_{\rm i}$ and fractional overlap $O_{\rm i}$, then {\bf lcmathrdbkg.f} assigns an uncertainty of $O_{\rm i}\times C_{\rm i}^{0.5}$, rather than  the Gaussian uncertainty $( O_{\rm i}\times C_{\rm_i})^{0.5}$ if no uncertainty is provided. Similarly, when an uncertainty $\sigma_{\rm i}$ is provided, then lcmath calculates the total uncertainty as $[ (O_1\times\sigma_1)^2 + (O_2\times\sigma_2)^2 + \dots ]^{0.5}$, when it should be $( O_1\times\sigma_1^2 + O_2\times\sigma_2^2 + \dots)^{0.5}$. Since $O_{\rm i}$ $<$ 1, $O_{\rm i}/O_{i}^{0.5}$ $<$1, and the uncertainties are underestimated.

 The underestimated Method I uncertainties are also responsible for the excess power in the PDS. This is because the Leahy PDS produced by {\bf powspec} are normalised such that the white noise level expected from the data errors corresponds to a power of 2 (see {\bf powspec} documentation). To do this, Gaussian errors are assumed and the normalisation, $A_{\rm Leahy}$, must be
\begin{equation}
A_{\rm Leahy} = 2\frac{\Delta t}{\overline{\sigma^2} N}
\end{equation} 
 where $\Delta t$ is the sampling time, $N$ is the number of points and $\overline{\sigma^2}$ is the mean squared error \citep[see e.g. Appendix A of ][]{vgn03}. I.e., $A_{\rm Leahy}$ $\propto$ 1/$\overline{\sigma^{2}}$.
 Therefore, underestimating the Method I uncertainties by a factor of 1.1--1.4 will overestimate the PDS power by a factor of $\sim$1.2--2.

\subsection{ Suppression of high-frequency power}

 After correcting for the power excess described above, we find that the Method I PDS actually dips below the Poisson noise level at high frequencies. I.e., the high frequency variability is suppressed, causing the break in the PDS at frequency $\nu_{\rm b}$.

 For those M31 sources identified with broken power law PDS, $\nu_{\rm b}$ = 20$\pm$4--57.8$\pm$0.5 mHz, independent of the luminosity.
  We have studied the break frequencies of the 190,000 lightcurves that were constructed from three non-simultaneous, randomly-generated lightcurves and discussed in Sect.~\ref{3lc}. We found $\nu_{\rm b}$ to vary over $\sim$3--70 mHz, with no strong correlation between $\nu_{\rm b}$  and offset or intensity. While the break frequency is necessarily related to the bin size, this is relation is very weak also because $\nu_{\rm b}$ varies by a factor of $\sim$30 while the bin size is constant.

\section{Ultra-luminous X-ray sources in NGC 4559 and NGC 5408}
\label{ulx}

 We conducted a brief literature search for published PDS of combined EPIC lightcurves that may also be affected by this data-analysis  artefact. 
 We found a published broken power law PDS for NGC 4559 ULX7 \citep{crop04} and also for a ULX in NGC 5408 \citep{sor04}. We re-analysed XMM-Newton observations of these sources, first following the methods described in these papers, using non-synchronised lightcurves, then comparing their PDS with those from synchronised lightcurves. However, it is beyond the scope of this paper to re-analyse all published XMM-Newton variability data!

\begin{figure}[!b]
\resizebox{\hsize}{!}{\includegraphics[angle=270]{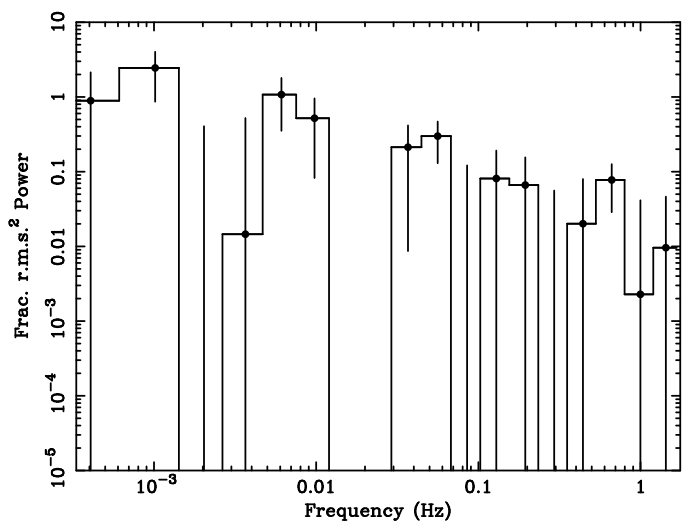}}
\caption{  Method II PDS of NGC 4559 ULX-7. The power is normalised to give fractional r.m.s.$^2$ variability. The expected noise  is subtracted, following \citet{crop04}. This PDS is consistent with zero power ($\chi^2$/dof = 20/16, with a null hypothesis probability of 0.22). Hence there is no evidence for the the variability reported by \citet{crop04}.
}\label{fig6a}
\end{figure}

\citet{crop04} reported a broken power law  PDS from an  ultra-luminous X-ray source (ULX) in NGC 4559 with a UV/X-ray luminosity exceeding 2.1$\times$10$^{40}$ erg s$^{-1}$.   We extracted MOS1, MOS2 and pn lightcurves from a source region with 30$''$ radius using Method II. We constructed a PDS that was normalised to give fractional r.m.s.$^2$ variability, and the expected noise was subtracted. The expected noise had a power of 4.3. The frequency bins were geometrically grouped.  The resulting PDS is shown in  Fig.~\ref{fig6a}. This PDS is consistent with zero power (good fit probability = 0.22); hence, the variability reported by Cropper et al. (2004) is artificial. 

 We note that \citet{fk05} also re-analysed the XMM-Newton observation of NGC 4559 ULX-7, and were able to reproduce the variability reported by \citet{crop04} only when the intervals of background flaring were included; hence, they mistakenly  attributed the variability to the background flares, unaware that \citet{crop04} also excluded the background flaring from their analysis. \citet{fk05} produced their PDS directly from the events lists that describe each photon (Kaaret 2006, private communication). Hence they did not observe the artefact after the intervals of high background were excluded.

\citet{sor04} studied a series of XMM-Newton observations of a ULX in the dwarf galaxy NGC 5408 over an 18 month period between 2001, July  and  2003, January. Its 0.2--12 keV spectrum requires two components, a power law with photon index $\Gamma$ = 2.6--2.9  and a thermal component with blackbody temperature of 0.12--0.14 keV; such a spectrum is typical of a black hole X-ray binary in the very high / steep power law state \citep{mr03}.
 The inferred 0.2--12 keV luminosity is $\sim$10$^{40}$ erg s$^{-1}$ \citep{sor04}. They present a broken power law PDS from $\sim$3 ks of the 2003, January XMM-Newton observation of NGC 5408, where $\gamma$ increases from $\sim$0 to 1.3$\pm$0.2 at 2.5 mHz. Such a PDS is also consistent with a black hole X-ray binary in the very high state. 

\begin{figure}[!b]
\resizebox{\hsize}{!}{\includegraphics[angle=270]{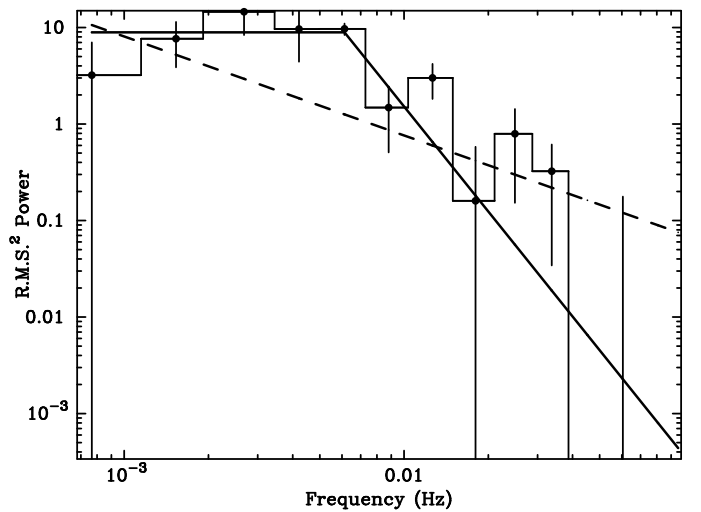}}
\caption{ Method II PDS for the NGC 5408 ULX, averaged over 3 intervals of 256 bins with 0.1 s resolution. The PDS is normalised to give the fractional r.m.s$^2$ variability, and the white noise is subtracted. Power law and broken power law fits are shown, with best fit $\chi^2$/dof of 77/12 and 24/11 respectively.
}\label{fig8}
\end{figure}

 We extracted EPIC lightcurves  from a circular source region with 30$''$, following \citet{sor04}. A Method II PDS is presented in Fig.~\ref{fig8}; the PDS is r.m.s$^2$ normalised, and averaged over 3 intervals of 256 bins, with 5 s time bins. Fitting this PDS with zero power results in a $\chi^2/dof$ of 92/14, hence the variability is real. Figure~\ref{fig8} also shows the best  power law and broken power law fits to the PDS, with $\chi^2$/dof = 77/12 and 24/11 respectively. The broken power law fit is clearly preferred (f-testing yields a 99.95\% chance that the improvement is significant),  but we caution that it is not an acceptable fit: the goodness of fit probability is just 1.2\%. 

\section{Conclusions}
\label{conc}
Some of the authors have previously reported variability in XMM-Newton observations of M31 X-ray sources that are characterised by broken power law PDS, which are signatures of disc-accreting X-ray binaries \citep{bok03,bko04,will05}. However, this ``Type A'' variability observed in XMM-Newton observations of M31 is an artefact and may be entirely attributed to errors introduced by {\bf lcmath} when combining non-synchronised lightcurves. (Barnard et al., this issue). However, extragalactic analogues of the Galactic X-ray binaries should exhibit Type A and Type B behaviour; Type A variability should be observed in sufficiently bright X-ray sources.

 In  the survey of XMM-Newton observations of X-ray sources in M31, we found that the Type A variability was observed at lower luminosities than Type B variability, as expected for LMXBs. However, the most likely explanation for this observed behaviour is that the offset was smaller for the brighter sources; the start time for an XMM-Newton lightcurve is defined by the arrival time of the first event, so brighter sources will naturally have smaller offsets between CCDs. 

The published broken power law PDS of NGC 4559 ULX7 is also artificial. The PDS of the NGC 5408 ULX shows genuine intrinsic variability. Timing analysis of XMM-Newton observations of extra-galactic X-ray sources is indeed viable.  M31 X-ray sources have exhibited  pulsations \citep{osb01,tru05}; bursts \citep{ph05}; periodic dipping due to photoelectric absorption \citep{tru02,man04}, from a precessing disc in one case \citep{bfh06}, finally branch movement in a trimodal colour-intensity diagram reminiscent of Galactic Z-sources \citep{bko03}.

\section*{Acknowledgments}

Power density spectra were fitted using {\bf fitpowspec}, provided by P.J. Humphrey. R.B is funded by PPARC, while S.T. acknowledges support from NASA grant NAG5-12390.  We thank the anonymous referee for their hard work and  constructive comments.

\bibliographystyle{aa-package/bibtex/aa}
\bibliography{m31}

\end{document}